\documentclass[twocolumn,amsmath,amssymb]{revtex4}
\usepackage{epsfig}
\usepackage{graphicx}
\usepackage{dcolumn}
\usepackage{bm}
\usepackage{color,soul}

\begin{document}

\preprint{preprint}

\title{Revelation of Mott insulating state in layered honeycomb lattice Li$_2$RuO$_3$}

\author{Sakshi Bansal, Asif Ali, B. H. Reddy$^\dagger$ and Ravi Shankar Singh}
\email{rssingh@iiserb.ac.in}
\affiliation{Department of Physics, Indian Institute of Science Education and Research Bhopal, Bhopal Bypass Road, Bhauri, Bhopal - 462 066, INDIA}


\begin{abstract}

We investigate the role of electron correlation in the electronic structure of  honeycomb lattice Li$_2$RuO$_3$ using photoemission spectroscopy and band structure calculations. Monoclinic Li$_2$RuO$_3$ having Ru network as honeycomb lattice undergoes magneto-structural transition at T$_c$ $\sim$ 540 K from high temperature phase $C2/m$ to low temperature dimerized phase $P2_1/m$. Room temperature valence band photoemission spectra reveal an insulating ground state with no intensity at Fermi level ($E_F$). Ru 4$d$ band extracted from high and low photon energy valence band photoemission spectra reveal that the surface and bulk electronic structures are very similar in this system. Band structure calculations using generalized gradient approximation (GGA) leads to metallic ground state while screened hybrid (YS-PBE0) functional reveals opening up of a gap in almost degenerate $d_{zx}$/$d_{yz}$ orbital, whereas $d_{xy}$ orbital is already gapped. Ru 3$d$ core level spectra with prominent unscreened feature provides direct evidence of strong electron correlation among Ru 4$d$ electrons which is also manifested by $|E-E_F|^2$ dependence of spectral density of states (DOS) in the vicinity of $E_F$ in the high-resolution spectra, establishing Li$_2$RuO$_3$ as Mott insulator.

\end{abstract}


\maketitle

\section{INTRODUCTION}
Quasi two dimensional transition metal oxides (TMOs) with layered honeycomb lattice have recently been topic of great interest due to plethora of novel phases in these compounds \cite{K-H-model,QSHE-Ir,yogesh-Ir,Miura-struc,Miura-MO,VBLiquid} and their potential applications \cite{Li-ion1,Li-ion2}. In these systems, the transition metal at the center of the octahedra forms a layered honeycomb network. Among these, honeycomb iridates $A$$_2$IrO$_3$ ($A$ = Na, Li) have received considerable interest  due to anisotropic exchange interaction between spin and orbital moments in the strong spin-orbit coupling (SOC) limit, leading to novel ground states \cite{K-H-model,QSHE-Ir,yogesh-Ir}. Ru doped Li$_2$IrO$_3$ demonstrates that models based on the assumption of only SOC do not describe this system properly \cite{Ir-Ru}.  In addition to SOC, on-site electron correlation ($U$) among $d$ electrons is also expected to play an important role, as found in relativistic Mott insulator Na$_2$IrO$_3$ \cite{NIO-j1/2} as well as in many other transition metal oxides \cite {BJ-j1/2, cairo3, mgir-znir, rss-STIO}. On-site electron correlation and its role in deciding the novel ground state properties have been extensively studied in TMOs \cite{MIT} where the systems can be defined by a single parameter $U$/$W$ ($W$ = width of the $d$ band) within the Mott-Hubbard model and the Mott criteria $U$/$W$ $\sim$ 1 separates the Mott insulators ($U$/$W$ > 1) and correlated metals ($U$/$W$ < 1) \cite{Mott,Hubbard}. 3$d$ TMOs fall in the strongly correlated regime due to the narrow 3$d$ band. Since, $U$ inversely depends on the extension of the radial wave function of the $d$ orbital, correlation effects are expected to be less important in wide band 4$d$ and further in 5$d$ TMOs \cite{RuO2,casrruo3,IrO2,bairo3}. Surprisingly, moderate to large correlation effect has been realized in various 4$d$ and 5$d$ systems exhibiting bad metallic to Mott insulating ground states \cite{y2ru2o7,ca2ruo4,ru-core,Ca2FeReO6,rss-YIO,sr3iro7}.

4$d$ TMOs, particularly Ru-based oxides, exhibit a plethora of interesting properties such as superconductivity, non-Fermi liquid behaviour, unusual magnetic ground states, {\em etc.}, while exhibiting varying electron correlation strength \cite{Sr2RuO4,kalo-epl,rana-sci-rep}. 4$d$ analogue of the honeycomb iridates is Li$_2$RuO$_3$ which crystallizes in monoclinic structure with Ru network forming honeycomb lattice. It undergoes first order magneto-structural transition at $\sim$ 540 K from high temperature phase $C2/m$ to low temperature dimerized phase $P2_1/m$ accompanied by a steep increase in resistivity (insulating to highly insulating behaviour) and a steep decrease in magnetic susceptibility with the decrease in temperature \cite{Miura-struc}. In the dimerized phase, strong disproportionation of about 19\% between short and long Ru-Ru bonds has been observed at room temperature. Dimerization occurs due to strong overlap between one pair of Ru 4$d$ orbitals resulting into formation of molecular orbitals triggering the structural transition below 540 K \cite{Miura-struc,Miura-MO,yogesh-Ru}. Li$_2$RuO$_3$, with partially filled $t_{2g}^4$ band, is expected to be metallic. In contrast, transport measurements reveal it to be highly insulating suggests correlation effects may also be important, as also emphasized in cluster dynamical mean field theory (LDA+cDMFT) calculations \cite{dmft}.

In this letter, we investigate the role of electron correlation in the electronic structure of  dimerized Li$_2$RuO$_3$ using photoemission spectroscopy and band structure calculations. Valence band collected using $x$-ray and ultra-violet photoemission spectroscopy reveal that the surface and bulk electronic structures are very similar. Ru 4$d$ band exhibits a broad peak at around 1 eV binding energy along with a shoulder at higher binding energy. No intensity at the Fermi level confirms the insulating ground state. Band structure calculations using YS-PBE0 functional opens up a gap in correlated $d_{yz}$/$d_{zx}$ orbitals whereas $d_{xy}$ orbital is already gapped in GGA calculation, revealing the orbital dependency of correlation effect in this system. Prominent unscreened features in Ru 3$d$ core level spectra and parabolic energy dependence of spectral DOS near $E_F$ suggest the influence of strong electron correlation among Ru 4$d$ electrons.

\section{EXPERIMENTAL AND CALCULATION DETAILS}

Polycrystalline samples of Li$_2$RuO$_3$ were synthesized by solid-state reaction method using high purity ingredients Li$_2$CO$_3$ (99.995\%) and RuO$_2$ (99.99\%). Stoichiometric amounts of ingredients were mixed, and the well ground mixture was pressed into pellets and calcined at 700 $^o$C for 12 hours. To achieve large grain size, samples in highly pressed pellet form was sintered at 1000 $^o$C for 6 days with two intermediate grindings. Samples were furnace cooled at the end of each heat treatment. A 10\% excess of Li$_2$CO$_3$ was used to compensate for the evaporation of Li during heat treatment. The phase purity was confirmed by powder $x$-ray diffraction (XRD) pattern collected at room temperature using PANalytical X'Pert Pro diffractometer equipped with Cu $K\alpha$ radiation ($\lambda$ = 1.540 \AA). Rietveld refinement of $x$-ray diffraction pattern was performed using the HighScore Plus software. XRD pattern shows no secondary phase including the absence of any peak corresponding to RuO$_2$. The Rietveld refinement reveals that the crystal structure is monoclinic with lattice parameters $a$ = 4.933 \AA, $b$ = 8.778 \AA, $c$ = 5.894 \AA, and $\beta$ = 124.41$^o$ and crystallizes in $P2_1/m$ space group which are in excellent agreement with earlier reports \cite{Miura-struc}. Photoemission spectra were collected at room temperature on {\em in-situ} (base pressure $\sim$ 4 $\times$ 10$^{-11}$ mbar) fractured samples using spectrometer equipped with a Scienta-R4000 electron energy analyzer with a total instrumental resolution of 400 meV, 8 meV and 5 meV for monochromatic Al $K\alpha$ (1486.6 eV), He {\scriptsize II} (40.8 eV) and He {\scriptsize I} (21.2 eV) photons (energy), respectively. Multiple samples were fractured to ensure the reproducibility of data and cleanliness of the surface was ensured by the negligibly small feature at higher binding energy in O 1$s$ core level spectra and absence of feature corresponding to C 1$s$ core level. Polycrystalline silver was used to determine $E_F$ and the energy resolutions for different photon energies at 30 K.

Band structure calculations were performed using full-potential linearized augmented plane wave (FPLAPW) method as implemented in Wien2k for experimentally found structural parameters consisting of 4 formula units (fu) in the unit cell \cite{wien2k}. 17$\times$8$\times$14 $k$ mesh within the first Brillouin zone and GGA \cite{GGA} were used to calculate the DOS. To compare with the experimental valence band and to take care of the electron correlation in the band structure calculations, we also performed the calculations using YS-PBE0 functional \cite{YS-PBE0} and DOS was calculated for 10$\times$4$\times$8 $k$ mesh within the first Brillouin zone. The local coordinate system for Ru atom, where $x$ and $y$ axis points to the oxygen atoms on the common edge shared by two RuO$_6$ octahedra (forming the Ru-Ru dimer), is chosen for obtaining orbital resolved DOS for Ru 4$d$ states. For all the calculations, the energy and charge convergence was better than 0.1 meV and 10$^{-4}$ electronic charge, respectively.

\begin{figure}[tb]
	\centering
	\includegraphics[width=1.1\textwidth,natwidth=1200,natheight=750]{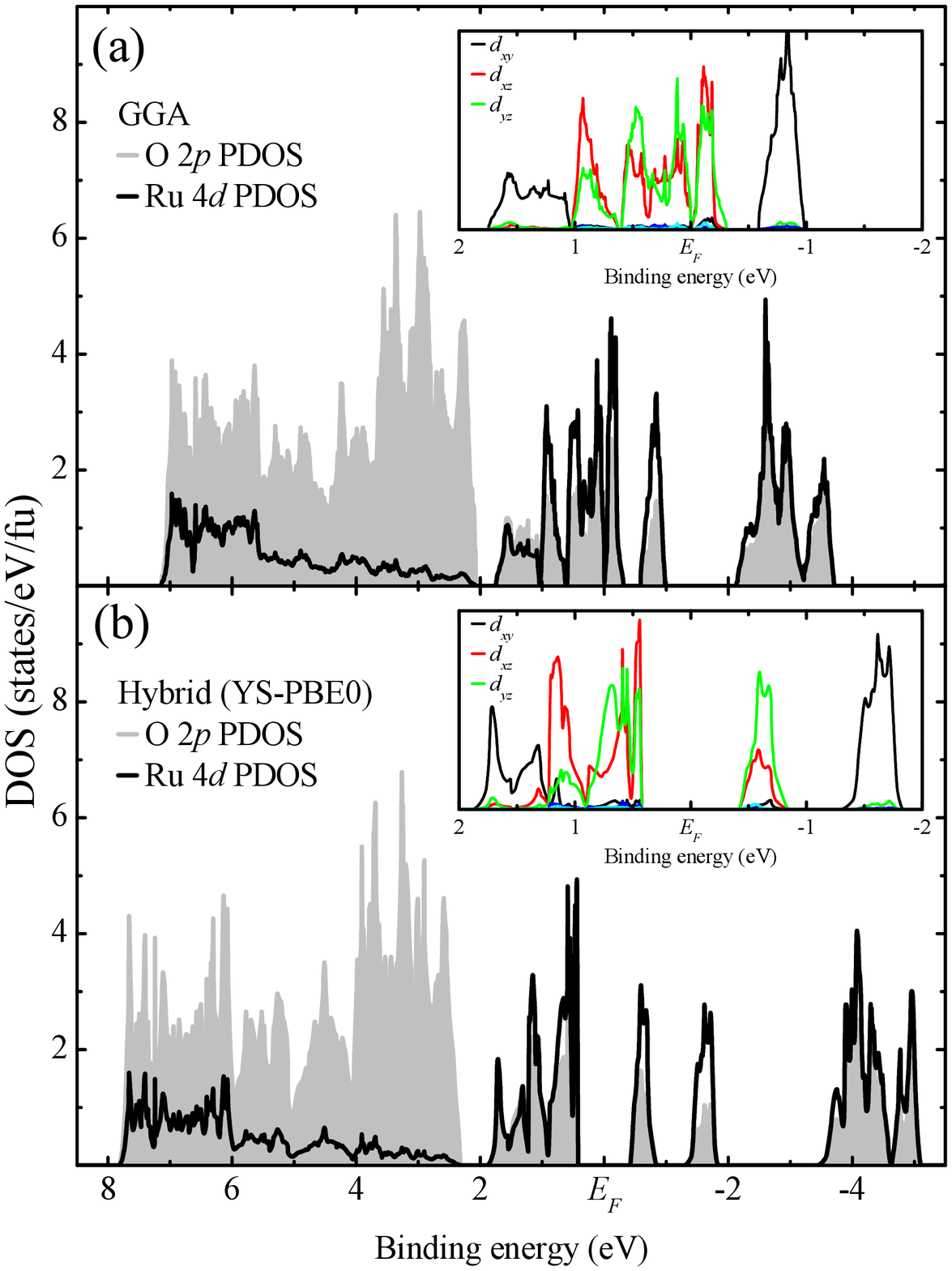}
	\vspace{-18ex}
	\caption{\label{fig:epsart} O 2$p$ and Ru 4$d$ PDOS for Li$_2$RuO$_3$ from (a) GGA calculations and (b) YS-PBE0 calculations. Insets in respective figures show different $d$ orbital contributions in $t_{2g}$ band in the local coordinate system (see text).}
\end{figure}

\begin{figure}[tb]
	\centering
	\includegraphics[width=1.1\textwidth,natwidth=1200,natheight=750]{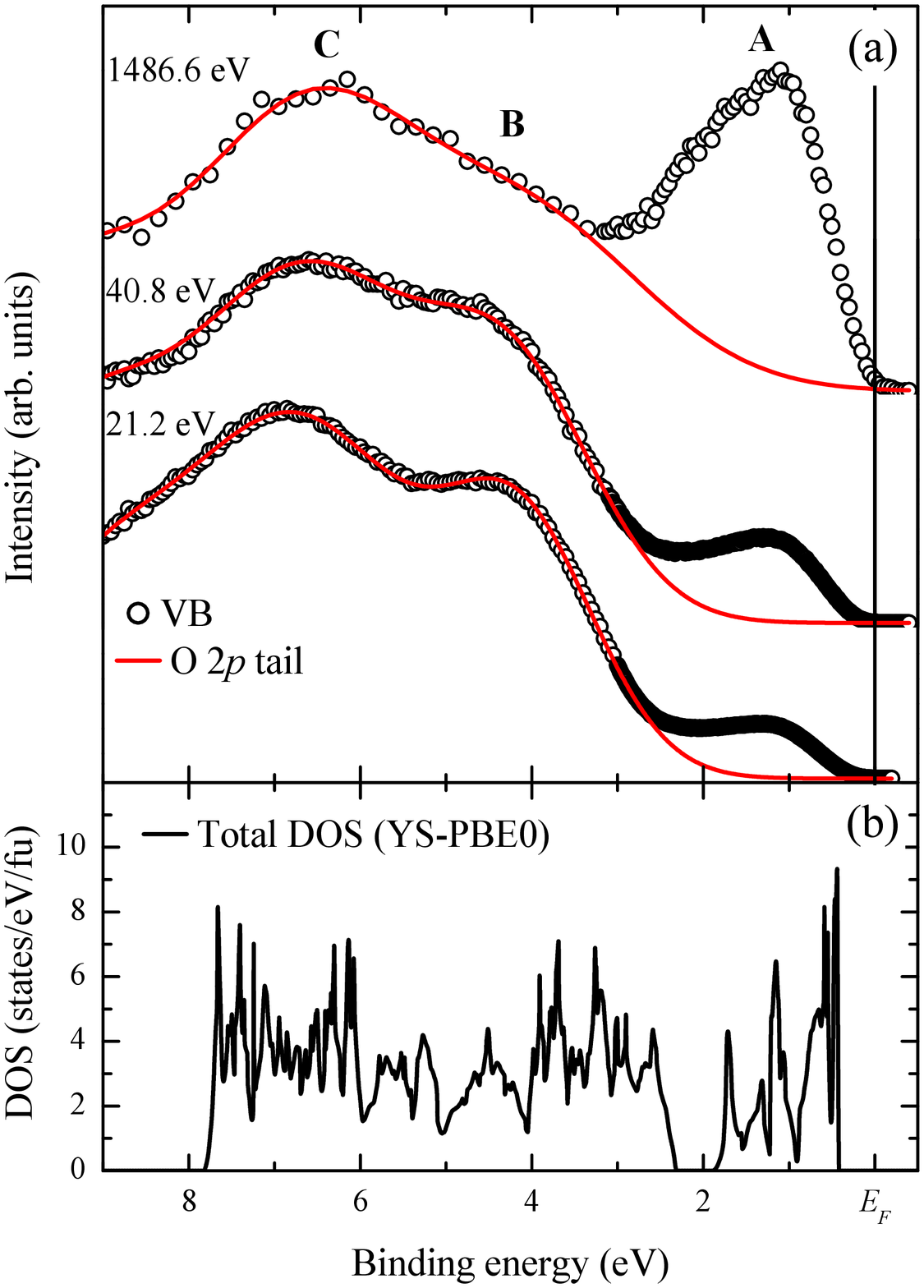}
	\vspace{-13ex}
	\caption{\label{fig:epsart} (a) XP, He {\scriptsize II} and He {\scriptsize I} valence band spectra of Li$_2$RuO$_3$ at room temperature. Lines show the O 2$p$ band contribution. (b) Total DOS for Li$_2$RuO$_3$ using YS-PBE0 calculations.}
\end{figure}

\section{RESULTS AND DISCUSSION}

In Fig. 1(a), we show the results of GGA calculations. Grey shaded area shows the O 2$p$ partial density of states (PDOS), and Ru 4$d$ PDOS has been shown with black line. PDOS corresponding to Li has not been shown here due to negligibly small contribution; thus, the valence band is formed by the hybridization of O 2$p$ with Ru 4$d$ states only. Three distinct groups of features are evident in the occupied part. Nonbonding O 2$p$ states appear between 2–5.5 eV, and the bonding states with dominant O 2$p$ contributions appear at higher binding energies (centered at $\sim$6 eV). The feature below 2 eV binding energy are antibonding states primarily having Ru 4$d$ character. In local coordinate system, $e_g$ band, comprising of $d_z^2$ and $d_{x^2-y^2}$ orbitals, is completely empty and appears below -2 eV binding energy while $t_{2g}$ band contribute in the vicinity of $E_F$ appearing between 2 to -1 eV binding energy. Total width of $t_{2g}$ states is about 2.7 eV and of $e_g$ states is about 1.5 eV. A dip like structure is observed, but no hard gap is found at $E_F$. The inset of Fig. 1(a) shows various $d$ orbitals of $t_{2g}$ band. It is to note here that $d_{xy}$ orbital is gapped and almost degenerate $d_{yz}$ and $d_{zx}$ orbitals are partially filled. The ideal Ru-honeycomb lattice (with perfect RuO$_6$ octahedra) exhibits $C_3$ symmetry, where $d_{xy}$, $d_{yz}$ and $d_{zx}$ orbitals are expected to be perfectly degenerate. While in case of distorted honeycomb lattice (low temperature dimerized phase), one of the Ru-Ru bond length has reduced, thereby breaking the $C_3$ symmetry and stronger direct overlap of the orbital along the dimer ($d_{xy}$ in the present case) leads to the formation of molecular orbital \cite{Miura-MO,dmft}.

Fig. 1(b) shows the PDOS corresponding to O 2$p$ and Ru 4$d$ states from band structure calculations using YS-PBE0 functional. Screened hybrid functional have been shown to be very successful in capturing the band gap in correlated systems where the ground states are often found to be semiconducting or insulating \cite{YS-PBE0,YS-PBE0-1,CSPO}. These results exhibit an insulating ground state with the gap opening in the $t_{2g}$ band; thus the $E_F$ has been set in the middle of the gap. This Mott gap has also been observed in the calculation using TB-mBJ \cite{anisotropy} and LDA+cDMFT \cite{dmft} calculations. The value of the energy gap of 0.84 eV is somewhat larger than other calculations \cite{anisotropy} and as well as the activation gap ($\sim$0.3 eV) from transport measurements \cite{gap}. The O 2$p$ band moves towards higher binding energy ($\sim$0.5 eV) in comparison to GGA results. The inset of Fig. 1(b) shows orbital resolved $t_{2g}$ band. Interestingly, strong electron correlation leads to opening of the gap in partially filled $d_{yz}$ and $d_{zx}$ orbitals having a small effect on $d_{xy}$ orbital (already gapped in GGA), as also seen in the LDA+cDMFT results \cite{dmft}. These results manifest the orbital selectivity of the correlation effects in Li$_2$RuO$_3$ where $d_{yz}$ and $d_{zx}$ orbitals are strongly correlated atomic orbitals and $d_{xy}$ forms the molecular orbital. Single site DMFT (LDA+sDMFT) calculations required quite large value of $U/W$ ($\sim$ 2.2) to exhibit the Mott gap in honeycomb lattice \cite{jafari} which later was shown to be $U/W$ $\sim$ 1 using LDA+cDMFT calculations. While the LDA+cDMFT has also been successful in describing the ground state of dimerized VO$_2$ \cite{VO2}, it is interesting to observe the Mott gap along with the orbital selectivity of correlation effects in dimerized Li$_2$RuO$_3$ using screened hybrid calculations.

Room temperature valence band spectra obtained using 1486.6 eV (XP spectra), 40.8 eV (He {\scriptsize II} spectra) and 21.2 eV (He {\scriptsize I} spectra) excitation energies are shown in Fig. 2(a). Similar to GGA and YS-PBE0 results, three discernible features A, B, and C are observed in the valence band spectra. The features B and C corresponding to O 2$p$ states are enhanced in the low excitation energy spectra, while feature A corresponding to Ru 4$d$ states is enhanced in XP spectra. This is due to the strong dependence of the relative transition matrix elements on excitation energies. The ratio of photoemission cross-section of Ru 4$d$ states to O 2$p$ states is significantly higher in XP spectra compared to that in He {\scriptsize II} and He {\scriptsize I} spectra \cite{yeh}.

The overall comparison of experimental spectra with GGA results in Fig. 1(a) suggests that a rigid shift of O 2$p$ band from GGA results ($\sim$0.5 eV towards higher binding energy) is required to match the observed peak position and widths of features B and C of the experimental spectra. Such a shift of completely filled O 2$p$ bands is often observed due to underestimation of correlation effects in the band structure calculations \cite{dd}.To compare the experimental results and to take care of the electron correlations in the band structure calculations we show the occupied part of the total DOS obtained using YS-PBE0 functional in Fig. 2(b). The total width of the O 2$p$ band as well as positions of the features A, B, and C are remarkably similar to the experimental spectra.

\begin{figure}[tb]
	\centering
	\includegraphics[width=1.1\textwidth,natwidth=1200,natheight=750]{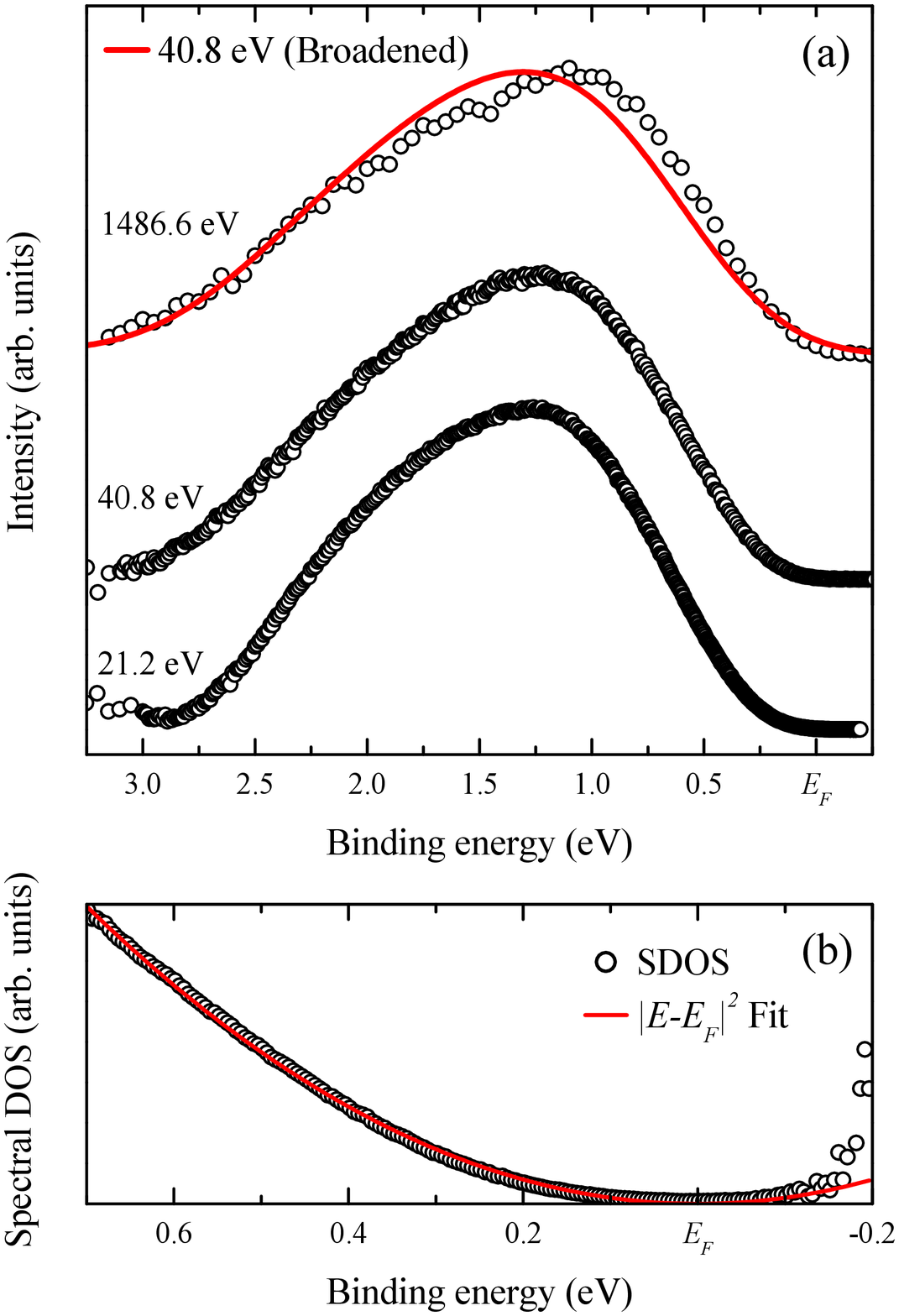}
	\vspace{-8ex}
	\caption{\label{fig:epsart} (a) Ru 4$d$ band extracted from XP, He {\scriptsize II} and He {\scriptsize I} valence band spectra. The resolution broadened He {\scriptsize II} spectra is shown by line. (b) Spectral DOS corresponding to He {\scriptsize I} spectra. $|E-E_F|^2$ fit is shown by line.}
\end{figure}

\begin{figure}[tb]
	\centering
	\includegraphics[width=1.1\textwidth,natwidth=1200,natheight=750]{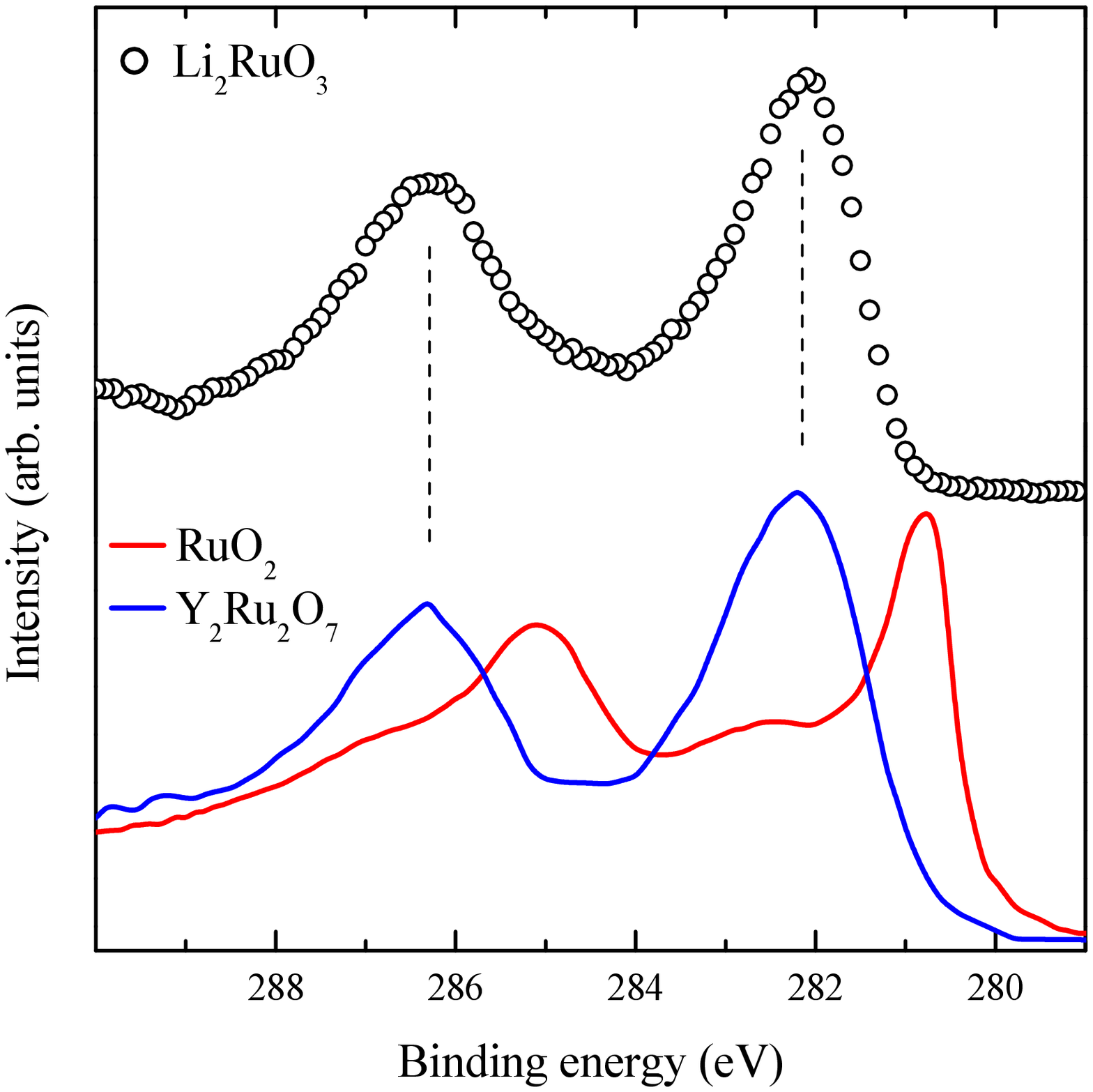}
	\vspace{-38ex}
	\caption{\label{fig:epsart} Ru 3$d$ core level spectra of Li$_2$RuO$_3$ at room temperature. Red and blue curves are corresponding to RuO$_2$ \cite{RuO2} and Y$_2$Ru$_2$O$_7$ \cite{y2ru2o7} respectively.}
\end{figure}

It is clear from Fig. 2(a) that the O 2$p$ and Ru 4$d$ bands are distinctly separated in high and low excitation energy spectra. Thus, Ru 4$d$ contributions can reliably be delineated by fitting O 2$p$ bands using two Gaussians representing feature B and C as performed in other systems \cite{casrruo3,bairo3,rss-YIO,rss-STIO}. Lines in Fig. 2(a) show the resultant fit obtained by the least-squares error method and the extracted Ru 4$d$ band is shown in Fig. 3(a). All the spectra, normalized by the total integrated intensity, show the main feature at around 1 eV binding energy, with a shoulder structure appearing between 1.5-2 eV binding energy. No intensity at the Fermi level in all the spectra suggests insulating character. Interestingly, the spectral line shape is very similar in both high excitation energy XP spectra as well as low excitation energy (He {\scriptsize I} and He {\scriptsize II}) spectra, despite having large difference in probing depth. The major difference in the low and high excitation energy spectra arises due to the energy resolution; thus, the resolution broadened (Gaussian broadening of 0.4 eV) He {\scriptsize II} spectra (shown by line) is compared with XP spectra. The very similar lineshape of these two spectra establishes that the surface and bulk electronic structures are essentially similar in contrast to the observations in other 4$d$ and 5$d$ transition metal oxides \cite{casrruo3,rss-YIO}. The spectral DOS can be obtained by dividing photoemission spectra with Fermi Dirac distribution function since the hole and electron lifetime broadening around Fermi level and energy broadening due to high resolution can be neglected. Thus obtained spectral DOS corresponding to He {\scriptsize I} spectra has been shown in Fig. 3(b). The line shows $|E-E_F|^2$ behavior of the spectral DOS in the vicinity of $E_F$.

It is to note here that lithium deficiency in Li$_2$RuO$_3$  (hole doping) would lead to larger DOS at $E_F$, which is reflected as well-screened feature in the Ru 3$d$ core level spectra  \cite{sakshi-AIP}. Increased DOS at $E_F$ is also observed in specific heat measurements for highly disordered Li$_2$RuO$_3$ \cite{gap}. The value of Sommerfeld coefficient $\gamma$, for the least disordered Li$_2$RuO$_3$ is similar to that of the band insulating Li$_2$TiO$_3$, suggests negligible DOS \cite{gap} as observed here in the high resolution He {\scriptsize I} spectra. The parabolic energy dependence of spectral DOS manifests strong correlation induced soft Coulomb gap at $E_F$ in this finitely disordered system \cite{rss-STIO, rss-YIO, Efros}.

In Fig. 4, we show Ru 3$d$ core-level spectra of Li$_2$RuO$_3$ collected at room temperature. The core level spectra exhibit spin-orbit split peaks at 282.2 eV and 286.3 eV corresponding to Ru 3$d_{5/2}$ and Ru 3$d_{3/2}$ with the spin-orbit splitting of about 4.1 eV. Absence of any signal at $\sim$280 eV binding energy in the present case of {\em in-situ} fractured sample confirms the high quality sample (with least disorder/lithium deficiency) \cite{sakshi-AIP}. For reference, we also show the Ru 3$d$ core level spectra for RuO$_2$ and Y$_2$Ru$_2$O$_7$. Ru 3$d$ core level spectra, in various ruthenates having Ru in 4+ valence state, exhibit two peak structures \cite{ru-core} for each spin-orbit split peak as seen in case of metallic RuO$_2$ \cite{RuO2}. The peak appearing at around 281 eV and 285 eV correspond to well-screened features while weak and broad features at around 282.5 eV and 286.5 eV correspond to poorly-screened features. These features are associated with the final state effects where a core hole generated in the photoemission process are screened by the electrons at the Fermi level. In contrast to metallic RuO$_2$, Mott insulating Y$_2$Ru$_2$O$_7$ only exhibits broad peaks corresponding to unscreened features since the valence electrons are localized in the presence of strong electron correlation leading to the disappearance of screened feature \cite{y2ru2o7}. The spectral features with respect to peak position and broadening in the case of Li$_2$RuO$_3$ are very similar to Y$_2$Ru$_2$O$_7$, confirming the strong correlation and thus the Mott insulating state in the present system. 

\section{CONCLUSION}
In conclusion, we have investigated the electronic structure of layered honeycomb lattice Li$_2$RuO$_3$ to understand the role of electron correlation on the electronic structure. Room temperature valence band spectra suggest an insulating ground state with no intensity at $E_F$. The line shape of Ru 4$d$ band extracted from high and low energy photoemission spectra, having different probing depths, suggests surface and bulk electronic structures are very similar in this system. The influence of electron correlation is manifested by parabolic energy dependence of spectral DOS in the vicinity of $E_F$. Band structure calculations using GGA and screened hybrid functional reveals orbital selective Mott state in this system. Ru 3$d$ core level spectra with prominent unscreened feature provide direct evidence for strong electron correlation among 4$d$ electrons. 

\section*{ACKNOWLEDGEMENTS}
We thank R. Kewat and N. Ganguly for the fruitful discussion. Authors acknowledge the support of Central Instrumentation Facility and HPC Facility at IISER Bhopal. Support from DST-FIST (Project No. SR/FST/PSI-195/2014(C)) is also thankfully acknowledged. \\

$^\dagger$ Present address: Department of Physics, Government College (A), Rajamundry – 533105, A. P., India

\end{document}